\pdfoutput=1
\documentclass{article}
\usepackage[hidelinks]{hyperref}
\usepackage{color}
\usepackage{graphicx}
\usepackage{comment}
\usepackage[backend=bibtex, style=numeric, bibencoding=ascii, maxnames=10]{biblatex}
\newcommand{\SBcomment}[1]{[#1]}      

\newcommand{\eg}{{\em e.g.}}

\begin{document}
\title{Dependability in Edge Computing}
\author{Paul Wood, Heng Zhang, Muhammad-Bilal Siddiqui, Saurabh Bagchi \\
School of Electrical and Computer Engineering \\
Purdue University}
\date{}
\maketitle


\section{Abstract}
\label{sec:abstract}

Edge computing is the practice of placing computing resources at the edges of the Internet in close proximity to devices and information sources. This, much like a cache on a CPU, increases bandwidth and reduces latency for applications but at a potential cost of dependability and capacity. This is because these edge devices are often not as well maintained, dependable, powerful, or robust as centralized server-class cloud resources\footnote{{\em Terminology}: We distinguish between two classes of devices---the client devices and the edge computing devices, or simply edge devices. When used without qualification, a device refers to a client device.}. This article explores dependability and deployment challenges in the field of edge computing, what aspects are solvable with today's technology, and what aspects call for new solutions.

The first issue addressed is failures, both hard (crash, hang, etc.) and soft (performance-related), and real-time constraint violation. In this domain, edge computing bolsters real-time system capacity through reduced end-to-end latency. However, much like cache misses, overloaded or malfunctioning edge computers can drive latency beyond tolerable limits.
Second, decentralized management and device tampering can lead to chain of trust and security or privacy violations. 
Authentication, access control, and distributed intrusion detection techniques have to be extended from current cloud deployments and need to be customized for the edge ecosystem.
The third issue deals with handling multi-tenancy in the typically resource-constrained edge devices and the need for standardization to allow for interoperability across vendor products. 




We explore the key challenges in each of these three broad issues as they relate to dependability of edge computing and then hypothesize about promising avenues of work in this area.

\section{What is different about Dependability in Edge Computing?}
\label{sec:definition}

For the purpose of this article, we will consider as edge devices, those that are in the premises of the end user (the home or the industrial campus) as well as those are outside the premises, say at the edge of the Internet
(\eg, content distribution nodes at the edge) or of the cellular network (\eg, base station). This definition is consistent with prior use of the term \cite{bonomi2012fog, satyanarayanan2017emergence}, though it is wider than some other prior usages \cite{hu2015mobile}.
\SBcomment{In terms of geographical spread of the coordinating edge devices, our definition of edge could span from a small number of edge devices deployed in a neighborhood, those deployed in cellular base station within a city, to a city-wide deployment.}
Local device-level computation is offloaded to nearby edge computing devices (foglets, cloudlets, etc.) whenever local processing is either inadequate or costly, or the computation relies on non-local information. 
For example, a long-enough voice snippet from a phone can be processed at a cell tower rather than on a local device (mobile edge computing), both saving battery and reducing end-to-end latency due to processing speed differences.
In contrast with traditional heavily centralized cloud computing, the edge computers act as a distributed computing infrastructure, providing increased bandwidth and reduced latency but with limited resources when compared with a central cloud.

The edge paradigm supports the large scale of the Internet of Things (IoT) devices, where real time data are generated based on interactions with the local environment. This complements more heavy-duty processing and analytics occurring at the cloud level.
This structure serves as the backbone for applications, such as augmented reality and home automation, which utilize complex information processing to analyze the local environment to support decision making. 
In the IoT domain, functional inputs and outputs are physically tied to geographically distributed sensors and actuators. 
If this data is processed in a central location, immense pressure will be placed on ``last mile'' networks, and cloud-leveraged IoT deployments will become impractical. 
Without drastic network improvements, which seem unlikely in the mid-range future \cite{broadband2014state}, edge computing is likely to become a cornerstone of IoT.
We see that there has been a shifting of the envelope of local versus edge computing, based on two dimensions---first, as more demanding applications arise (voice processing to video processing to augmented reality) and second, as the locally available resources increase (processing, storage, networking). 
The first drives some processing toward the edge (and further, toward the cloud) while the latter drives processing to move closer to local devices. 
In the context of edge computing dependability, we focus on the five aspects that we deem most significant: large scale, low latency or soft real-time requirements, authentication and physical security, multi-tenancy on the edge devices, and standardization.

\begin{figure*}
\label{fig:overview}
\centering
\includegraphics[width=\linewidth]{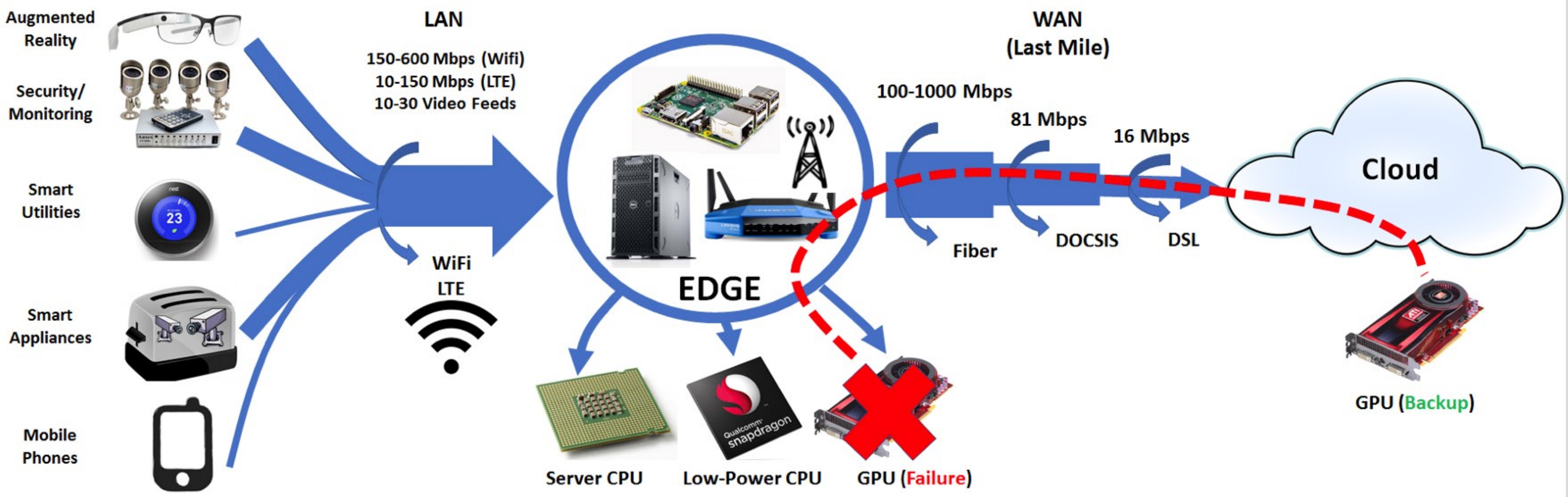}
\caption{High-bandwidth edge services on the left generate dense video or less dense environmental data that needs to be processed on the edge. The high bandwidth of WiFi allows edge computers such as ARM-based Raspberry Pi's, Routers, and traditional tower servers to process the data. When there is a failure, the data can be re-routed to the cloud, but this re-routing is limited by the capacity of the WAN system.}
\end{figure*}


\section{Resiliency Challenges}
\label{sec:challenges}

Applications that benefit from edge computing typically have requirements for low latency and 
generate high bandwidth data streams. Two canonical examples are provided by IoT devices and Augmented Reality (AR) applications. With this model, we now examine the resiliency challenges that are posed by edge computing applications. 

\subsection{Large Scale}
Since edge computing applications are in their infancy, many current design decisions seem reasonable at a small scale.
However, 
many practical challenges arise as the scale of edge applications grow, both in terms of the number of client devices and the amount of data being generated by them.
The Internet of Things,
propelled by low-cost wireless electronics and ease of integration, is greatly increasing the number of addressable computers and the amount of environmental data available for transmission on the Internet. 
Shared computing utilities, especially networking resources, can quickly become saturated by scale-out of data-intensive applications, and protocols can fail to deliver results in a timely manner when algorithmic complexity is super-linear in terms of the number of endpoints. 

\subsubsection{Network Impact}
Cloud-based computing supports scalability by incrementally adding resources to the computing environment as new devices enter service.
Practically, scale of this nature is more easily achieved in centralized datacenters where new server crates can be parked and connected to existing infrastructure.
With increasing scale of IoT, offloading all processing to the datacenter becomes infeasible since network operators rely on average case capacity for deployment planning, and network technologies have not kept up with the growth of data.
A fixed video sensor may generate 6 Mbps of video 24/7, thus producing nearly 2 TB of data per month--an amount unsustainable according to business practices for consumer connections, \eg, Comcast's data cap is at 1 TB/month and Verizon Wireless throttles traffic over 26 GB/month.
For example, with DOCSIS 3.0, a widely deployed cable-internet technology, most US-based cable systems deployed today support a maximum of 81 Mbps aggregated over 500 home---just 0.16 Mbps per home.
If 12 users are interacting with AR in the area, the network would be saturated.
With edge computing, all 500 homes could be active simultaneously but at a resilience cost: if each home has 4 active AR users, then only 3 of the 500 edge computers could fail simultaneously and fall back to cloud-based processing, due to the constraints of the last-mile network. 

\subsubsection{Lack of Failover Options}
The first resiliency challenge becomes a lack of resources to fail over to in case of a failure.
In the cloud, individual resource availability becomes less significant due to the presence of hot spares in the same local network.
In an edge environment, hot spares are not practical---a common deployment scenario is very lean with a single edge device, such as, a WiFi router providing all the edge services in a home. 
An alternative solution to this is to negotiate peer-based fail-over or community aggregation of resources (similar to microgrids).
In this case, the community bandwidth resources (of the 500 homes) will still limit failover capacity.
Without additional network infrastructure support, the 500 home community can only support 12 AR users that are not processing data locally, even if the failover peer is in the same local community.
With more bandwidth, additional redundant edge computers would become feasible, but this requires last-mile network support that is both expensive and has historically been slow to deploy.

\subsection{Failing to Meet Real-Time Deadlines}
The promise of low latency from edge computing attracts application deployment with soft real-time requirements.
AR applications, for example, need to remain below 16 ms end-to-end latency to maintain seamless 60 frame per second user interactions.
Such latency is easily achievable on local devices, but having {\em all} client devices have the requisite computing capability is infeasible.
Edge computing provides a cost saving aspect, especially if the edge device is idle most of the time (such as, a ``computer'' in the cash register of a store).
This leads to real-time applications operating on edge computers instead of on client devices.

Inside an edge computing device, a finite amount of CPU, RAM, GPU/APU, and networking support exist.
Each real-time application needs some slice of these resources to perform its task within the prescribed deadline.
Real-time scheduling in a constrained environment---a problem often solved by earliest deadline first---can be complicated by the intermix of delay tolerance levels from different applications, unpredictable user interactions, unpredictable network behavior between the edge device and the client device,
as well as drift in the clocks of multiple client devices being served by a common edge device~\cite{koo2009tale}.
Devices may wait to operate until capabilities can be secured from the edge computing network.
As an example, consider a video stream from an AR device being analyzed at an edge device and a thermostat now needs analysis of the video stream to determine how many people are in a room and adjust its setting accordingly. If the edge device can only support the analysis of a single video stream due to the demanding nature of such processing, it will have to make a scheduling decision to prioritize one of the two streams.
If the edge computing paradigm becomes popular, this almost guarantees contention for the resources at the edge devices, and the assumption of instant availability of resources that many applications make today, may cause failures of timeliness guarantees for many real-time services. One possible solution approach here is to use the cloud as a failure backup, for delay-tolerant applications. Also, where the application is stateful, and in view of the impermanence of some edge devices, the state may be stored on the client, on the cloud, or on a combination of the two.

\begin{table}
\centering
\caption{Edge Computer Costs}
\label{tbl:edgecosts}
\begin{tabular}{lllll}
               & Raspberry Pi & Router & Xeon E3-1220L & Specialized \\
Video Feeds & 1            & 0            & 1         & 1-4    \\
GFlops         & 1-2          & 0.5-1        & 50        & 100-200    \\
Cost           & \$50         & \$200        & \$300-500   & \$300-500  
\end{tabular}
\end{table}



\subsection{Authentication and Physical Security}
Edge computing needs to address security challenges especially considering that the client devices may be embedded in private physical spaces. 
In a nod to the anticipated large scale of these systems, the security mechanisms themselves need to be scalable and decentralized.
\SBcomment{We expect that the importance of the scalability concern will vary with the scale of the edge device deployment, from where it is local to a neighborhood (less of a concern) to where it is city-wide (more of a concern).}
We expect that economic imperatives will mean that most edge devices will be cheap. This will mean that any security mechanism that requires expensive hardware or has large memory footprint will be infeasible except for a small subset of edge devices. This opens up the design space of security mechanisms in a heterogeneous environment with a large number of constrained devices and a few (security) resource-rich devices. 

\subsubsection{Scalable Authentication}
Since client devices are placed close to information sources, they are necessarily distributed such that physical access to these devices cannot be protected. 
An attacker can do invasive probing and install malicious software on these devices.
As a result, any cryptographic keys stored on the device are subject to tampering and eavesdropping. 
These can be made more difficult by hardening methods, but they cannot be eliminated altogether.
Consequently, the authentication and trustworthiness of client devices must be validated through existing low-cost hardware security techniques such as code signatures, etc.
These techniques mostly rely on some form of public key infrastructure (PKI) which has a somewhat high computational cost, but most importantly, a high management cost \cite{butler2000national}.
The PKI systems may become cumbersome for the low-cost, high-volume OEMs of either client or edge devices to properly implement and manufacturers may opt out of secure system designs in favor of ad-hoc or proprietary mechanisms.

In this model, each device needs a managed key-based authentication system, whereby devices must be marked, signed, and managed after creation.
End users must be able to easily identify a device by its public information and verify through the OEM that the device is secure (\eg, has an untampered software stack) and properly authenticated before sharing information with the edge infrastructure. 
PKI and the SSL system used today for secure banking and other services can scale out to the IoT level, but the level of interaction changes.
In these traditional systems, a set of root public keys are distributed by operating systems to the end devices on a regular basis.
The end device must verify any secure host's published key against this set of root public keys. In this system, the scale of secure hosts is on the order of the number of public facing websites, software development companies, etc.
In an IoT scenario, the number of secure host certificates will scale to the billions, placing immense pressure on device manufacturers to manage the process of issuing, storing, and securing certificates and client and edge device keys.
Key management itself is often a weak point in both scalability and security \cite{pki-vulnerability-arstechnica-2012}, such as, keeping root keys safe on the devices or generating keys with enough entropy. 

In order to alleviate the concerns raised by public key systems, biometric authentication can be introduced in a home environment. Central device within a home can be authenticated biometrically and then the authentication can be propagated to all connected devices. As an extra layer of security, different devices can be mapped to different expiration times depending on the functionality of the device. For critical devices such as a locker or heart monitor, the expiration times can be very low and should be authenticated every time just before their use. Multiple expiration times combined with delegation of authentication from the central control in a house leads to scalable secure methods of information transactions. 

\subsubsection{Decentralized Security}
The fact that the edge network may be disconnected, or be in degraded connectivity, means that security mechanisms in the edge devices should be autonomous
and be capable of receiving on-demand updates, again in a secure manner \cite{bagchi2012protocol}. 
Because of high computing power involved in PKI systems, a low end edge device will need to delegate its computational tasks to a more powerful device nearby. For a time-critical operation, the edge device may not be able to establish connection and verify some operation with a central server that is connected to the Internet. We can take some inspiration from schemes for self-organized public key management for mobile ad hoc networks \cite{capkun2003self}, which also deal with disconnected and decentralized operation, but we will need closer attention to the latency involved. It may be possible to use Physical Unclonable Functions (PUF) to verify authenticity of the edge devices, without needing to contact a central data source.
Hence, in a geo-distributed and mobile environment, with issues ranging from intermittent connectivity with centralized infrastructure to complete connection outages, it should be possible to make security decisions autonomously, perhaps suboptimal from an efficiency or functionality standpoint but still meeting the privacy or security guarantees. Thus, it is important to design the equivalent of fail-safe modes of failure, analogous to what exists in safety critical systems. 
 
Along with providing authentication and infrastructure for security in a highly scalable edge ecosystem, we also need to develop decentralized security mechanisms tailored for the edge devices. 
Increasing concerns in privacy and data ownership, and impaired solutions from centralized cloud infrastructures warrants distributed, peer-to-peer security mechanisms 
to be implemented in the edge ecosystem that also eliminates the need of centralized privacy mediators. 
We will have to explore the secure distributed mechanisms such as blockchain that eliminates the need of centralized infrastructure as well as protects data privacy from the owners of any centralized infrastructure. Blockchain enables secure distributed peer-to-peer transaction exchange and makes these edge devices autonomously secure. In a home, a private blockchain environment can be created consisting of digital updates and secure data sharing between smart devices. Recently, industry has started to realize the potential of blockchains in the edge ecosystem \cite{ibm-blockchain}. Data from heterogeneous devices in home can be converted into blockchain ledger format through custom API and then used securely in a home monitoring system.

\subsection{Multi-Tenancy of Services and Billing}
An edge device, much like a computing node in the cloud, will need to support multiple tenants.
Clouds perform this by virtualization combined with a per-user public billing system. 
It has been found in the virtualization literature \cite{novakovic2013deepdive, maji2014mitigating} that while some resources can be well partitioned (like processing cores), some others are notoriously difficult to partition (like cache capacity and memory bandwidth), which leads to performance interference. 
In edge computing, since low latency is an important driver, it will be particularly important to prevent performance failures. 
Further, edge computers will need a similar billing system to properly manage resources in a congested system. 
For example, a smart thermostat from company A and a smart oven from company B may both wish to use edge computing resources. 
The billing question will depend critically on whether the edge device is within the premises and under a single ownership, or outside the premises and under shared ownership or providing service to multiple unrelated users. 
In the first case, the billing question is distinctly easier. 
Regarding multi-tenancy, deciding which client application gets what portion of the resource at the edge device will require input about ownership and applications' timing requirements.
Consumers may be unable to understand the nuances of contention and therefore an automated solution is necessary to solve this multi-objective optimization problem.
Popular virtualization technologies such as virtual machine, or even containers, may be too heavy-weight for edge devices. 
These need a relatively significant amount of hardware resources to execute, \eg, VMWare's ESX hypervisor needs a recommended 8GB of RAM, while a Docker container with NAT enabled doubles the latency of a UDP stream \cite{felter2015updated} (see Figure 3). 
Thus the challenge will be to find a lighter weight solution for multi-tenancy, possibly at the expense of reducing the isolation among the different applications and limiting the total number of applications. 


\subsection{Standardization}
In our daily life, we are increasingly seeing a proliferation of ``smart devices''.
However, these are being built essentially without regard to standardization and thus interoperability.
We expect that for edge computing to flourish, this trend will need to be arrested and instead, standardization put in place. 
Thus, new standards or new mediation layers should be designed to coordinate those devices to provide useful functionality, such that an edge device can seamlessly communicate with, and possibly control, multiple end user devices, fail-over from one device to another is possible. 

It will be economically viable to realize edge computing if relationships and risks among all parties can be clearly delineated. 
In the area of cloud computing, many standards have been proposed, such as NIST Cloud Computing Reference Architecture (CCRA) 2.0, ISO/IEC standard under the group ``Cloud Computing and Distributed Platforms'', and Cloud Standards Customer Council (CSCC) standard under ``Data Residency Challenges''.
These are slowly beginning to gain momentum, with some early signs of convergence toward the NIST standards. 
Likewise, interoperability in the edge computing landscape will require standardization for various aspects---visualization, data management, programming APIs, etc. 


However, existing standards from cloud computing cannot be copied over directly to edge computing due to many factors. 
One primary factor is the sensitivity of information being available at edge devices, such as more personal data being collected at finer time granularity and this increases the risks of exposure of sensitive data.
Another distinction is the service delivery requirement. 
In cloud computing, usually the computing task is more heavy duty with sizable amount of data transfer whereas in edge computing, there is more frequent and lighter communication. 
Therefore, the standard for edge computing must address the issues of service delivery latency and bandwidth differently.
Thus, it appears to us that a great deal of effort needs to be expended in standardization, with partnership among industry and non-profits, to create a flourishing edge computing marketplace.

\section{Application Challenges: A Smart Toaster}
\label{sec:future}


In this section, we illustrate edge computing's resiliency challenges inside of an example application. 
The challenges encompass practical aspects of integrating data-intensive applications (here, video-based controls) in an edge environment (here, the home) with low network bandwidth to the Internet.

\begin{figure}
\centering
\includegraphics[width=1\columnwidth]{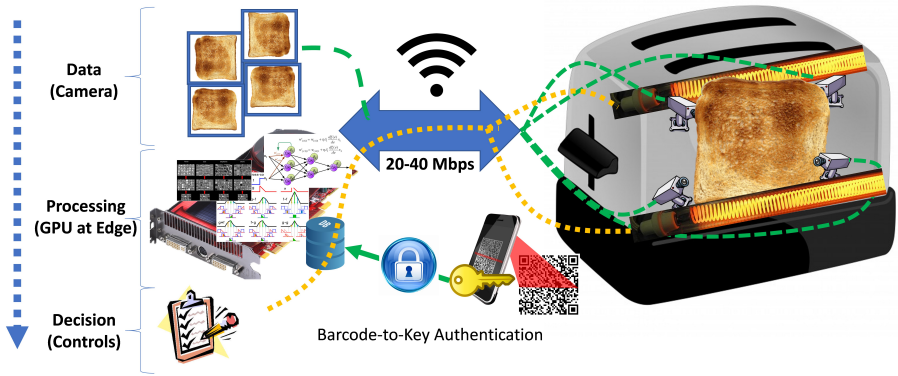}
\caption{Our smart toaster, a hypothetical example that carries the ordinary function of toasting bread to an absurd level of IoT-edge-integration to highlight the future research issues. It can toast various kinds of bread according to user preferences, using video processing to determine when the appropriate level of toasting has been reached. It can also order the right amounts of the right kind of bread. It relies on edge devices and optionally cloud processing to achieve its functionality.}
\label{fig:smart-toaster}
\end{figure}

We motivate our discussions with a hypothetical example: a smart toaster (Figure \ref{fig:smart-toaster}).
Our smart toaster takes a simple, ordinary function---toasting bread---to an absurd level of IoT-edge-integration to highlight the future research issues.
The device is a WiFi-connected toaster with video feeds of the toasting area, a motorized bread lift, and electronic heater controls. The toasting can be scheduled for a future time period and the appropriate number of breads can be retrieved from its storage area. 
All types of breads, from bagels to texas toast, are accepted, and the device classifies the bread based on its weight and video-derived characteristics.
The image processing utilizes a deep learning approach (\eg, CNN) that learns from the raw video feeds.
There are decisions made prior to toasting (what breads to schedule at what time based on prior user preferences) and some during toasting.

A typical workflow is that the scheduler schedules the right number of breads for toasting.
When the time is reached, the breads are inserted into the toaster. 
The video feed is processed by the CNN and a classification is created.
The toasting begins, and live video is streamed to the CNN to determine appropriate toast level.
Once the proper level is reached, as inferred by the video processing software, the toast is ejected from the oven area and a notification is sent to the user.
In the theme of our earlier discussions, the video bandwidth is 24 Mbps from four cameras at 6 Mbps each (1080p video). Delay tolerance in this domain is on the order of a second, however many edge computing applications will have more demanding delay requirements, such as augmented reality. Such differing delay tolerance requirements should be handled by the edge device, which will typically support multiple client applications. 

\subsection{Availability}
Latency-sensitive applications need graceful degradation support for operation under edge failure. 
For example, a smart toaster may utilize CNN-based processing for raw video feeds to determine proper toast level.
Under ideal circumstances, this video feed is sent to and processed at the local edge computing node (\eg, Google Home).
Some delay is tolerated, but a baking process cannot be interrupted and still achieve consistency.
If the CNN processor at the edge becomes unavailable due to a fault, then the local device must operate without such decision support.
Several solutions may exist:

\paragraph{Pre-Processed Decisions}
An acceptable toast time may be generated based on prior toasting events. 
Whenever the process starts, the toaster is programmed with a fail-safe decision about when to stop.
This approach represents classic average-case static control systems.
For example, most toast needs 5 minutes of oven time, so the toasting system is setup for that default.

\paragraph{Supervisory Control}
Running a full CNN may not be practical at the edge device when it is under contention from other client applications, but alternative low-power algorithms can exist simultaneously with the CNN logic.
If the processing time exceeds the delay tolerance level, or is predicted to, then a switch can be made to the alternative, simpler processing block. 
Mean pixel color, for example, can serve as a lower intensity processing alternative to determine adequate toastiness.
This represents more robust but still simple control algorithms that can reliably provide improvements even under contention situations.

\paragraph{Degraded Operational Modes}
Trusty toaster controls have existed for quite some time, and often users can accept a temporary lack of feature availability as long as it is not too disruptive. 
Manual toasting controls as an override can provide degraded mission support, but the user loses the ability to schedule toasting {\em a priori}.
When high bandwidth connection to the cloud is available, it may be possible to determine the availability of the bread for ordering from nearby retailers. When such connection is unavailable, a degraded mode of operation could be to simply inform the user of the fact that the supply in the toaster storage has run out. The general principle is that the edge computing applications must be designed with multiple degraded modes of operation in mind. 

\subsection{Resource Contention}
Since edge computing enters the domain of real-time control, the edge resources must be properly managed to avoid contention issues.
Similar problems arise in distributed smart grids--overloading the grid even temporarily can cause issues.
Some accepted solutions rely on constraint-managing dispatch systems that classify different loads by their requirements.
For example, in \cite{petersen2013taxonomy}, power loads are classified as batteries, bakeries, and buckets (BBB). A bakery is the kind of load where the process must run in one continuous stretch at constant power consumption. The Bakery could be a commercial green house, where plants must receive a specific amount of light each day. This light must, however, be delivered continuously to stimulate the photosynthesis of the plants. Our toaster example is this kind of load.
In the edge computing scenario, each client application registers with a resource manager, and the devices can effectively reserve resources prior to execution.
In the case of the toaster, this means that the CNN process and bandwidth for the video feed are reserved prior to starting the toast process.
This thrust indicates that there is research to be done for the appropriate level of reservation and scheduling under time constraints. We can rely on significant prior work in the area of soft real-time systems. Two domain specific challenges however arise here. First, the delay tolerance can be specific to the context, \eg, the specific user using the device. This needs to be programmed in, and in the longer term, learned by the scheduler for the edge resource. Second, there are several levels of resources available for making the scheduling decision---the client device, the edge device, and resources on the cloud. Each choice has interdependent effect on choices made for other client applications.

\subsection{Authentication}
Authentication will exist in two pieces. 
The first piece, credentials, will be relatively straight forward to solve.
As with SSL, a collection of central authorities (such as, Azure, EC2, Rackspace) will provide API's for registering and generating signed public/private key pairs from their IoT support systems.
The second piece, access control, must contain the association between users, their devices, and the edge computers.
This piece is complicated by both scale and usability factors.

We propose the best solution will be a distributed management layer for device-to-edge association.
Simple trusted interfaces can be established in the local domain, via physical access, and used as a gateway for additional device association.
Armed with a trusted root certificate, an edge device can verify the certificates of all peer devices locally. 
Once trusted, the identity simply needs to be added to the access control list for the local edge system.
The list can be managed by a trusted smartphone application with access to add a key, corresponding a new device, to the list.
In a simple case, each client device could have a barcode, and the application can be used to identify the device and its public key simply by scanning it with the phone's camera.
The key requirement would be to simplify the user involvement.


\section{Conclusion}
Edge computing presents an exciting new computational paradigm that supports growing geographically distributed data integration and data processing for the Internet of Things and augmented reality applications. 
Edge devices and edge computing interactions reduce network dependence and support low-latency, context-aware information processing in environments close to the client devices. 
However, services built around edge computing are likely to suffer from new failure modes, both hard failures (unavailability of certain resources) and soft failures (degraded availability of certain resources).
Low latency requirements combined with budget constraints will limit the fail-over options available in edge computing compared to a traditional cloud-based environment.
Consequently, system developers must develop and deploy applications on the edge with an understanding of such constraints.
Additional issues that edge computing will face include authentication at scale, cost amortization, and resource contention management. Further, for a thriving ecosystem, it is essential to have standardization of the device and network APIs, something that has not been seen to date. 
How these issues are handled will ultimately determine the success or failure of the paradigm of edge computing. Like many technology inflection points, timely moves on the thrusts outlined in this article can significantly tilt the balance in favor of success.

\section*{Sidebar: Current Edge Deployment}
An example of a current edge deployment, albeit not with the richness of factors described in this article is the 4G radio access network (RAN) that uses Mobile Edge Computing (MEC) to better deliver content and applications to end users \cite{mobile-edge-computing}. It can adapt the service delivery according to radio link load and avoid long distance transmission by using local content caching. MEC was approved as a formal specification by the European Telecommunications Standard Institute (ETSI) in 2014 and there is ongoing activity in standardizing it. Two popular current use cases with this edge technology are to provide localized video at a stadium or concert venue, and asset tracking in a large enterprise with enterprise small cell networks. 


\section*{Sidebar: Natural failure case}

The mission of a Tsunami Warning Centre (TWC) is to provide early warnings on potentially destructive tsunamis \cite{tsunami-warning-center}. A TWC uses local and global seismographic networks transmitting seismograms in real-time to continuously monitor seismic activity in order to locate and size potential tsunamis. There is a great need for speed here because tsunami waves can travel at the speed of a commercial jet plane, over 500 mph, in deep ocean waters. The response time of a TWC when it can rely on local data and processing is 2-5 minutes and this is increased by a factor of 10 when it needs to rely on distant data and processing [XXX]. Precisely due to earthquake or pre-earthquake activity, infrastructure may be disrupted, such as, power lines and communication lines, as has happened in multiple past tsunamis such as the 2011 Tohoku tsunami and Fukushima nuclear meltdown. Such disruption to infrastructure can impact the response time when there is reliance on distant data and processing. A resilient edge computing infrastructure can be properly harnessed and coordinated to collect data, do local processing, and aid in localized disaster mitigation efforts. Resilience implies that the factors that disrupted the current infrastructure should not affect the edge computing infrastructure. This is possible through localized communication infrastructure and the inherent redundancy in the edge devices. This is a particularly relevant use case because it satisfies the two key characteristics of edge computing, namely, geo-specific data and processing and requirement for low latency. 


\section*{Sidebar: Networking Challenges}
IoT and Edge Computing are positioning themselves to upend traditional consumer models for network usage.
Since the inception of a consumer-level Internet, content has existed in a central location, and consumers have downloaded that content, be it a website, audio file, or movie.
This near-constant trend has led to extensive asymmetric network deployments where finite bandwidth resources are partitioned to favor download over upload.
For example, the 2016 FCC definition of broadband\footnote{https://www.fcc.gov/reports-research/reports/broadband-progress-reports/2016-broadband-progress-report} is 25 Mbps download and only 3 Mbps upload, an 8:1 ratio.
Such ratios make sense for traditional Internet applications---users do not produce much content.
Slowly this has begun to change with the advent of live streaming video applications such as Periscope and YouTube Live, but even in these deployments, the viewer-to-stream ratio is still quite high.
IoT will upend this ratio, so that one Internet user (a home, for example) may produce 5+ video streams (as security cameras for example) for only 1 real-time viewer.
Given the current Internet needs, it will be impractical to scale down the 25 Mbps download to make room for the 25 Mbps upload, given finite radio spectrum and despite the frantic activity to release white space broadcast TV spectrum for communications needs.
Instead, more expensive technologies such as Google Fiber's active networks will have to replace aging infrastructure.
Since such deployments are slow and expensive and unlikely to gain universal penetration, and IoT devices are ready today, edge computing is a viable solution to the constrained network problem.


\begin{small}
\printbibliography
\end{small}

\end{document}